\documentclass{iaus}
\usepackage{graphicx}

\newcommand{\kms}{km~s$^{-1}$}

\title[Molecular Disks in Early-type Galaxies] 
{Disk Growth in Bulge-Dominated Galaxies: Molecular Gas and Morphological
Evolution}

\author[L.~Young et al.]   
{Lisa Young$^1$, Martin Bureau$^2$, Alison Crocker,$^2$ \break \and
 Francoise Combes$^3$}

\affiliation{$^1$Physics Department, New Mexico Tech, Socorro, NM 87801, USA
\break email: lyoung@physics.nmt.edu\\[\affilskip]
$^2$Sub-department of Astrophysics, University of Oxford, Keble Road, Oxford,
OX1 3RH, UK\\[\affilskip]
$^3$Obervatoire de Paris, LERMA, 61 Av. de l'Observatoire, 75014, Paris,
France}

\pubyear{2007}
\volume{245}  
\pagerange{xxx--xxx}
\date{?? and in revised form ??}
\jname{Formation and Evolution of Galaxy Bulges}
\editors{M. Bureau, E. Athanassoula, \& B. Barbuy, eds.}
\begin{document}

\maketitle

\begin{abstract}
Substantial numbers of morphologically regular early-type (elliptical
and lenticular) galaxies contain molecular gas, and the quantities of
gas are probably sufficient to explain recent estimates of the current
level of star formation activity. This gas can also be used as a
tracer of the processes that drive the evolution of early-type
galaxies. For example, in most cases the gas is forming dynamically
cold stellar disks with sizes in the range of hundreds of pc to more
than one kpc, although there is typically only $1\%$ of the total
stellar mass currently available to form young stars. The numbers are
still small, but the molecular kinematics indicate that some of the
gas probably originated from internal stellar mass loss while some was
acquired from outside. Future studies will help to quantify the role
of molecular gas (dissipational processes) in the formation of
early-type galaxies and their evolution along the red sequence.
\end{abstract}

\firstsection 

\section{The Molecular Gas Content of Early-type Galaxies} 

About a decade ago, surveys for molecular gas in early-type galaxies
began to show significant detection rates (e.g.\ \cite[Lees et al.\
1991]{lees91}; \cite[Wiklind et al.\ 1995]{WCH95}; \cite[Knapp \&
Rupen 1996]{knapp96}). The difficulty was that the surveys had very
strong selection biases towards galaxies that are bright at 100$\mu$m,
and it was virtually impossible to know how to extrapolate to
FIR-faint galaxies. In recent years this limitation has been removed
and there are now deeper surveys of less biased samples. \cite{ws03}
found a surprisingly high CO detection rate of $78\%$ in a
volume-limited sample of nearby field lenticular galaxies, and
\cite{swy07} detected CO emission in $33\%$ of a similar sample of
field ellipticals. \cite{cyb07} also detected CO emission in $28\%$ of
the early-types in the SAURON survey (\cite[de Zeeuw et
al.~2002]{dezeeuw02}), which is not a complete sample in any sense but has the
advantage that the gas content can be compared to information on the
stellar populations, internal dynamics, and ionized gas. In short,
when the samples are not FIR-biased there are still a significant
number of early-type galaxies showing CO emission. Cold gas masses are
highly variable however, with $M_{\rm gas}/L_B$ in the range $10^{-1}$
to $10^{-3}$ and lower.


The strongest connection between cold gas and galaxy properties is as
expected with indicators of recent star formation, including stronger
H$\beta$ emission and absorption (\cite[Combes et al.\
2007]{cyb07}). \cite{lucero07} also discuss the connections between
molecular gas, star formation, radio continuum and FIR emission, and
the CO detection rate in early-types is at least superficially
consistent with the incidence of star formation derived from GALEX UV
data (e.g.\ \cite[Yi et al.\ 2007]{yi05}; \cite[Kaviraj et al.\
2007]{kaviraj06}). Thus, the present gas content of early-type
galaxies does clearly reveal the potential for star formation and disk
growth.

\section{Disk Growth}

The morphological changes that can be effected in early-type galaxies
by star formation, and the ease with which feedback can disrupt it,
will all depend on the distribution of the molecular gas. A few
early-types have now had their CO emission mapped by millimeter
interferometers, e.g.\ \cite{inoue96}, \cite{wiklind97}, \cite[Young
(2002, 2005)]{young2002}, \cite{okuda05}, \cite{das05}, and
\cite{ybc08}. The molecular gas is typically found in regular,
kpc-scale rotating disks with masses of a few $10^{7}$ to a few
$10^9$~M$_\odot$ of molecular hydrogen. The disk radii range from a
tenth up to a few effective radii of the stellar distribution.  The
average surface densities are often $100$ M$_\odot$~pc$^{-2}$ and
higher, comparable to the molecular surface densities in spirals.

\begin{figure}
\includegraphics[scale=0.35]{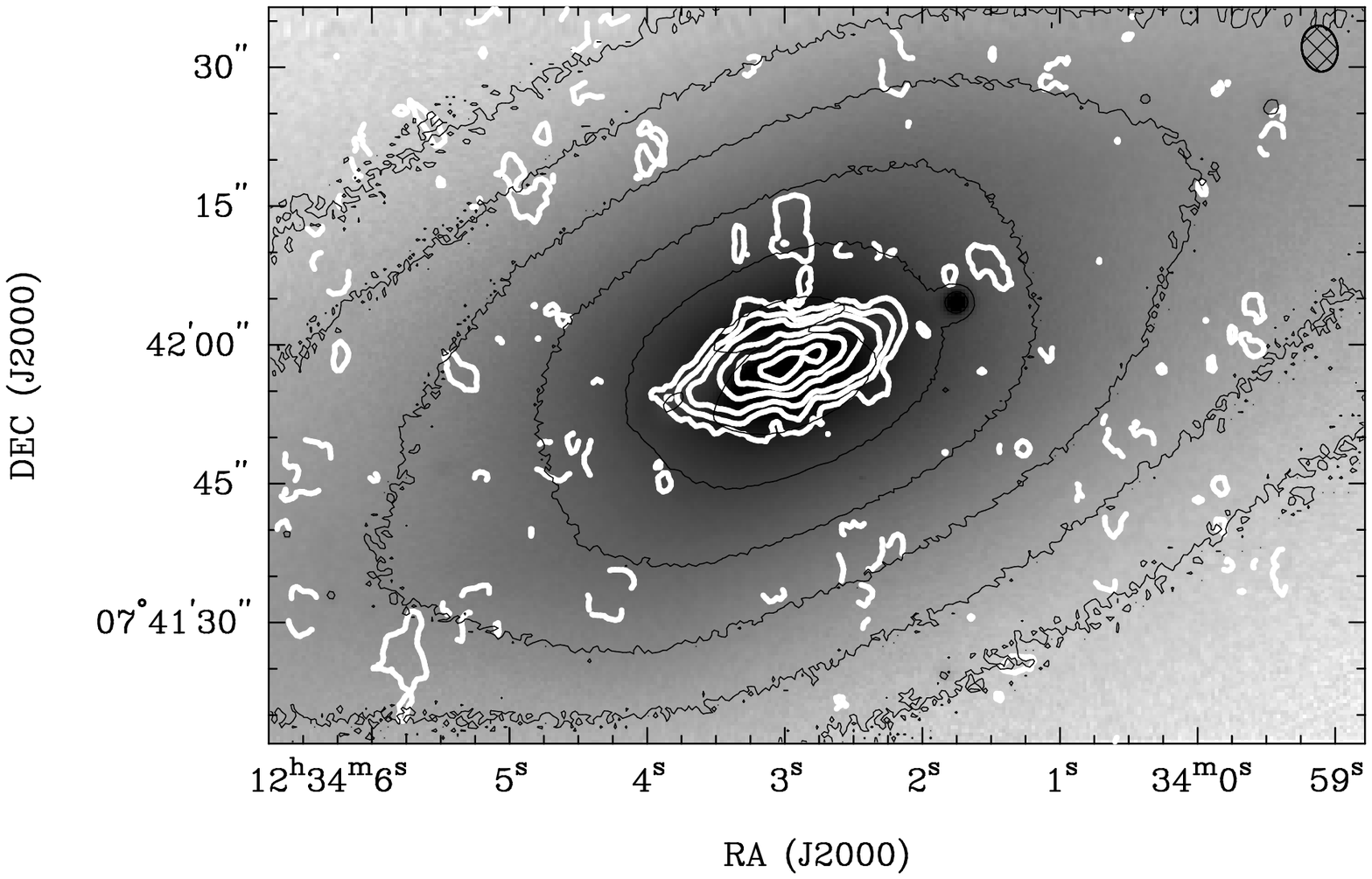}\hspace*{6mm}\includegraphics[scale=0.35]{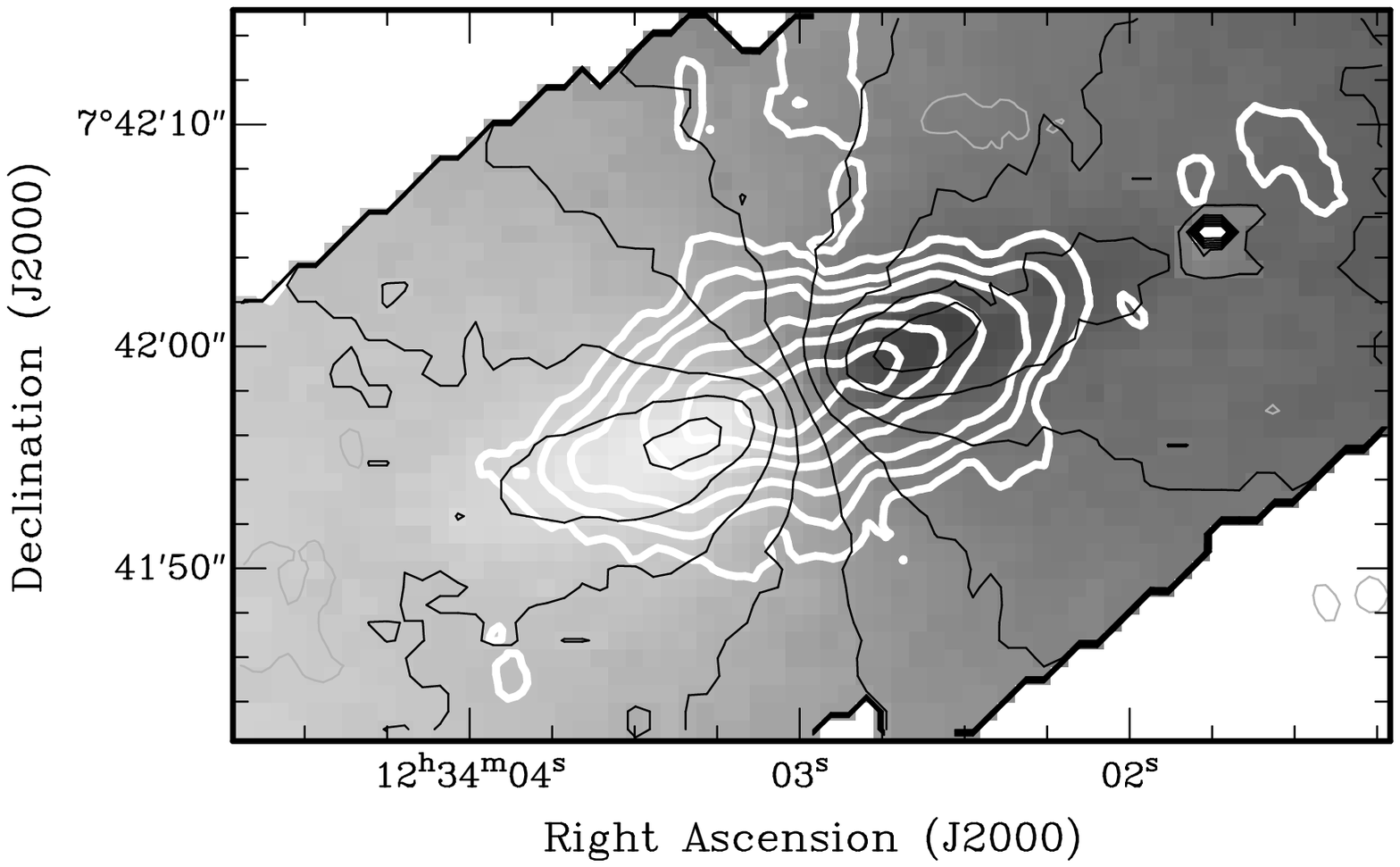}
\caption{Integrated CO intensity in NGC~4526 (white contours),
  overlaid on respectively a SDSS $g$ image (left) and the stellar
  velocity field (right; greyscale and black contours; \cite[Young et
  al.\ 2008]{ybc08}). CO contours are at $-10$, $10$, $20$, $30$,
  $50$, $70$, and $90\%$ of the peak
  ($37$~Jy~beam$^{-1}$~km~s$^{-1}$). Velocity contours are from $416$
  to $816$ \kms\ in intervals of $50$ \kms.}
\label{fig:ngc4526}
\end{figure}

When both molecular maps and two-dimensional optical spectroscopy are
available, there is usually good evidence for a young stellar disk
coincident with the molecular disk. The most dramatic example is the
Virgo cluster lenticular NGC~4526 (see Fig.~\ref{fig:ngc4526};
\cite[Young et al.\ 2008]{ybc08}), which has embedded in its bulge a
dynamically cold kpc-scale stellar disk with a substantially younger
mean age than the rest of the bulge (\cite[Emsellem et al.\
2004]{emsellem04}; \cite[Kuntschner et al.\ 2006]{kuntschner06}). The
[O III]/H$\beta$ line ratios are also suggestive of current star
formation activity (\cite[Sarzi et al.\ 2006]{sarzi06}) and the
ionized gas kinematics follow the molecular kinematics, albeit at
somewhat smaller velocity amplitude. Based on the FIR flux, the star
formation rate in NGC~4526 is roughly $1$ M$_\odot$~yr$^{-1}$, giving
a gas depletion timescale of $7\times10^8$ yr.  But the total
molecular mass now present in the galaxy is only $0.6\%$ of the total
stellar mass, so that even if there was substantially more molecular
gas in the recent past it can not make more than a small frosting of
young stars. The Virgo lenticular NGC~4459 is a similar case
(\cite[Young et al.\ 2008]{ybc08}).

\section{The Origin of the Cold Gas}

For at least 30 years it has been speculated that early-type galaxies
should have a substantial cold ISM originating in mass loss from their
own evolved stars (\cite[Faber \& Gallagher 1976]{FG76}). A modern
version of the theory, complete with heating, cooling, a proper
cosmological context and the influence of AGN can be found in
\cite{temi07} and references therein. It is also well known that
mergers and the accretion of satellites and/or cold gas can be
important mechanisms for galaxy growth. The relative importance of
internal and external gas sources for early-type galaxies is not yet
well known, but angular momentum can provide a strong constraint since
internally produced gas must have a specific angular momentum
consistent with that of the stars in the galaxy.

\cite{sarzi06} have measured the distribution of the misalignment
angle between the stellar and ionized-gas kinematic major axes for the
SAURON sample. They find evidence for ionized gas from both internal origins
(a large peak of galaxies with no misalignment) and external origins (a long
tail of misalignments out to $180^\circ$, counterrotation).
The corresponding data for molecular gas are much more rare, but
appear to be consistent with the above picture (see also \cite[Bureau
\& Chung 2006]{bureau06}). Of the four lenticulars mapped by
\cite{ybc08}, one (NGC~3032) has molecular gas counterrotating with
respect to the stars, so this gas must be left over from a major
merger or acquired from an external source. In the obvious merger
remnant NGC~3656, the gas disk is roughly perpendicular to the stellar
kinematic major axis (\cite[Young 2002]{young2002}), and in NGC~2768
the gas forms a polar ring (or disk) structure (\cite[Crocker et al.\
2008]{cbyc08}), but much more commonly the gas disks are aligned with
the mean stellar rotation and suggest a contribution from internal
mass loss. NGC~4150 harbours a counter-rotating young stellar core,
but strangely the CO gas is rotating like the galaxy bulk (\cite[Young
et al.\ 2008]{ybc08}).

The misaligned molecular disks in NGC~3032 and in NGC~3656 are both
massive, kpc-scale, and vigorously forming stars. Clearly, even when
the molecular gas has an external origin, it does not always descend
into the nucleus of the galaxy to feed and/or be destroyed by an
active nucleus. A better global understanding of the influence of the
accreted cold gas on early-type galaxies therefore requires an
accurate model for the impact parameters of this gas and the size
distribution of the resulting gas disks.

\section{Open Questions}


We have seen that kpc-scale molecular disks with $10^8$ to $10^9$
M$_\odot$ of molecular gas are relatively common in early-type
galaxies at $z=0$, and the incidence of these disks is probably
consistent with current star formation rate estimates based on UV
data. Environmental effects on the gas content of early-type galaxies
have not been properly explored yet, nor have the possible effects of
AGN activity to clear cold gas. The observed molecular disks
clearly document the process of disk growth within bulges, a process
which must have been of greater importance in the past, but the
details remain sketchy.

It is not yet known whether the molecular gas always forms stars (and
whether that result is consistent with our theoretical understanding
of star formation processes). It has been hypothesized that molecular
disks should be stabilized by the galaxies' steep gravitational
potentials (e.g.\ \cite[Kennicutt 1989]{kenn89}; \cite[Okuda et al.\
2005]{okuda05}; \cite[Kawata et al.\ 2007]{kawata07}). Clearly our
data have shown that the molecular gas is often associated with
current star formation, but perhaps this is not always the case. In
NGC~2768 the central molecular disk (only $\approx200$ pc) is located
in a prominent drop in the stellar velocity dispersion, just as one
would expect if a dynamically cold stellar disk had formed
(\cite[Crocker et al.\ 2008]{cbyc08}). Curiously, however, there is no
corresponding H$\beta$ absorption feature which would suggest current
or recent star formation. Early-type galaxies may thus present a
valuable testing ground for galaxy-scale theoretical models of star
formation.

Another open question has to do with the role of gas in the formation
of early-type galaxies in general, and more specifically in the origin
of the kinematically distinct substructures that are so common in
early-type galaxies (e.g.\ \cite[McDermid et al.\ 2006]{mcdermid06}).
The SAURON galaxies contain a large variety of cold stellar disks
(some of them metal-enriched) and misaligned or counterrotating cores,
in many sizes, and surely these are important clues to the galaxy
formation processes. It would be useful to know what fraction of the
substructures (or which kinds) are formed through dissipationless
processes and which required gas. The answer to this question will
likely require studies of more complete samples, such as Atlas$^{\rm
3D}$.

Nevertheless, very preliminary indications are that the molecular gas
probably originated in a combination of internal stellar mass loss and
cold accretion from external sources. Again, the available numbers are
still small, and a more quantitative assessment requires a substantial
increase in the number of galaxies with both cold gas and stellar
kinematic maps.

\begin{acknowledgments}
L.Y. is partially supported by NSF AST-0507432, and is indebted to all
the members of the galaxy group at the University of Oxford for
welcoming her into such a stimulating environment during her recent
sabbatical.
\end{acknowledgments}

%



\end{document}